\documentclass[12pt]{article}
\usepackage[english,german,french,polish]{babel}
\usepackage[T1]{fontenc}
\usepackage{amsfonts}

\begin{document}

\selectlanguage{english}

\baselineskip 0.73cm
\topmargin -0.4in
\oddsidemargin -0.1in

\let\ni=\noindent

\renewcommand{\thefootnote}{\fnsymbol{footnote}}

\newcommand{\SM}{Standard Model }

\newcommand{\SMo}{Standard-Model }

\pagestyle {plain}

\setcounter{page}{1}



~~~~~~
\pagestyle{empty}

\begin{flushright}
IFT-- 10/8
\end{flushright}

\vspace{0.4cm}

{\large\centerline{\bf Photoproduction of sterile scalars}}

{\large\centerline{\bf from cold dark matter through the photonic portal{\footnote{Work supported in part by Polish MNiSzW scientific research grant N N202 103838 (2010--2012).}}}}

\vspace{0.5cm}

{\centerline {\sc Wojciech Kr\'{o}likowski}}

\vspace{0.3cm}

{\centerline {\it Institute of Theoretical Physics, University of Warsaw}}

{\centerline {\it Ho\.{z}a 69, 00--681 Warszawa, ~Poland}}

\vspace{1.0cm}

{\centerline{\bf Abstract}}

\vspace{0.6cm}
  
\begin{small}


Operating with the recently proposed model of hidden sector of the Universe, based on the idea of photonic 
portal, we consider a new possible astrophysical process of weak  photoproduction of sterile scalars from the 
cold dark matter consisting of sterile Dirac fermions. Subsequently, these scalars weakly decay into secondary couples of photons. Then, primary photons, if energetic enough  and interacting in sufficiently dense dark-matter 
configurations, may form secondary-photon cascades appearing as a new astrophysical phenomenon. Primary 
photons, probing the cold dark matter, may come {\it e.g.} from gamma bursts. We show how the cross-section 
for the photoproduction and the rate for the decay can be calculated (in the centre-of-mass frame and at rest, respectively, for simplicity).
 
\vspace{1.0cm}

\ni PACS numbers: 95.30.Cq , 95.35.+d , 96.50.S- 

\vspace{1.0cm}

\ni August 2010

 
\end{small}

\vfill\eject

\pagestyle {plain}

\setcounter{page}{1}

\vspace{0.5cm}

\ni {\bf 1. Introduction}

\vspace{0.4cm} 

In a model proposed recently [1,2] there exists a hidden sector of the Universe, consisting of sterile spin-1/2 Dirac fermions ("\,$\!$sterinos"), sterile spin-0 bosons ("\,$\!$sterons") and sterile nongauge mediating bosons ("$\!A$ bosons")  described by an antisymmetric-tensor field (of dimension one) weakly coupled to antisterino-sterino and steron-photon pairs through the interaction energy

\begin{equation}
- \frac{1}{2} \sqrt{f}\left(\varphi F_{\mu \nu} + \zeta \bar\psi \sigma_{\mu \nu} \psi \right) A^{\mu \nu}\,.
\end{equation}

\vspace{0.2cm}    

\ni Here, $F_{\mu \nu} = \partial_\mu A_\nu - \partial_\nu A_\mu$ is the \SMo electromagnetic field (of dimension two), while $\sqrt{f}$ and $\sqrt{f} \zeta$ stand for two dimensionless small coupling constants. In the interaction (1), it is conjectured that the steron field $\varphi$ is broken into the sum 

\begin{equation}
\varphi = <\!\!\varphi\!\!>_{\rm vac}\! + \,\varphi_{\rm ph}
\end{equation}

\vspace{0.1cm}    

\ni  with  $<\!\!\varphi\!\!>_{\rm vac} \neq 0$ being a spontaneously generated vacuum expectation value of $\varphi$. The coupling (1) of photons to the hidden sector has been called "photonic portal"\, (to hidden sector) and it is an alternative to the popular "Higgs portal"\, [3].  

The new weak interaction Lagrangian (1), together with the $A$-boson kinetic and \SMo electromagnetic Lagrangians, provides the following field equations for $F_{\mu \nu}$ and $A_{\mu \nu}$:

\begin{equation}
\partial^\nu (F_{\mu \nu} +  \sqrt{\!f}\, \varphi A_{\mu \nu}) = -j_\mu \;\;,\;\; F_{\mu \nu} = \partial_\mu A_\nu - \partial_\nu A_\mu 
\end{equation}

\vspace{0.2cm}    

\ni and 

\begin{equation}
(\Box - M^2)A_{\mu \nu} = - \sqrt{f} (\varphi F_{\mu \nu} + \zeta \bar\psi \sigma_{\mu \nu} \psi) \,,
\end{equation}

\vspace{0.2cm}    

\ni where $j_\mu$ describes the \SMo electric current and $M$ stands for a mass scale of $A$ bosons, expected typically to be large. 

The field equations (3) are Maxwell's equations, the first of them modified in the presence of hidden sector (including sterons and $A$ bosons in addition to sterinos). Such a modification has a magnetic character, since the hidden-sector contribution to the total electric source-current

\begin{equation}
j_\mu + \partial^\nu ( \sqrt{\!f}\, \varphi A_{\mu \nu})
\end{equation}

\ni for the electromagnetic field $A_\mu$ is a four-divergence giving zero contribution to the total electric charge 

\vspace{0.1cm}    

\begin{equation}
\int d^3x[j_0 + \partial^k ( \sqrt{\!f}\, \varphi A_{0 k})] = \int d^3x j_0 = Q\,.
\end{equation}

\vspace{0.1cm}    

In particular, it can be seen that the vacuum expectation value $<\!\!\varphi\!\!>_{\rm vac} \neq 0$ generates spontaneously a small sterino magnetic moment

\begin{equation}
\mu_\psi = \frac{f \zeta}{M^2} <\!\!\varphi\!\!>_{\rm vac} \,, 
\end{equation}

\vspace{0.1cm}    

\ni though sterinos are electrically neutral and elementary. This is a consequence of an effective magnetic interaction


\begin{equation}
- \frac{1}{2} \mu_\psi \,\bar\psi \sigma_{\mu \nu} \psi\, F^{\mu \nu}  
\end{equation}

\vspace{0.1cm}    

\ni arising, when the low-momentum-transfer approximation     

\begin{equation}
A_{\mu \nu} \simeq \frac{\sqrt{f}}{M^2}\left(\varphi F_{\mu \nu} + \zeta \bar\psi \sigma_{\mu \nu} \psi \right) \,,
\end{equation}

\vspace{0.1cm}    

\ni implied by Eq. (4), is used in the interaction (1) with $\varphi = <\!\!\varphi\!\!>_{\rm vac}\! + \,\varphi_{\rm ph}$.

In our model, sterinos are stable and so are candidates for cold dark matter. In pairs, they can annihilate in a simple way into steron-photon pairs, $\bar{\psi}\psi \rightarrow \varphi_{\rm ph} \gamma $, and also into \SMo fermion pairs, $\bar{\psi}\psi \rightarrow \bar{f} f$ [1.2]. In contrast, sterons and $A$ bosons are unstable, decaying simply as $\varphi_{\rm ph} \rightarrow \gamma \gamma $ and $A \rightarrow \bar{f} f $. The decay $A \rightarrow \bar{f} f $ was discussed in Ref. [2]. The scattering of nucleons on sterinos ({\it i.e.}, on the cold dark matter), $\psi N \rightarrow \psi N $, was considered in Ref. [4].

In the present note, we calculate in the framework of our model the decay $\varphi_{\rm ph} \rightarrow \gamma \gamma $, but first, the photoproduction of sterons from sterinos ({\it i.e.}, from the cold dark matter), $\psi \gamma \rightarrow \psi \varphi_{\rm ph}$.

\vspace{0.4cm}    

\ni {\bf 2. Photoproduction of sterons from sterinos}

\vspace{0.4cm}

Consider the inelastic scattering of photons on sterinos with production  of physical sterons: $\psi \gamma \rightarrow \psi A^* A^*\varphi_{\rm ph} \rightarrow \psi \varphi_{\rm ph}$. This weak photoproduction process from cold dark matter is induced by a part of the weak coupling (1), 

\begin{equation}
- \frac{1}{2} \sqrt{f}\left(\varphi_{\rm ph} F_{\mu \nu} + \zeta \bar\psi \sigma_{\mu \nu} \psi \right) A^{\mu \nu}\,,
\end{equation}

\ni providing the following $S$-matrix element (in an obvious notation):

{\baselineskip 0.9cm

\begin{eqnarray}
S(\psi \gamma \rightarrow \psi \varphi_{\rm ph}) & = & -i\,\frac{1}{4}\,\frac{f \zeta }{k^2_A\!-\!M^2} \left[
\frac{1}{(2\pi)^{12}}\, \frac{m^2_\psi}{4E' E \omega_\varphi \omega} \right]^{1/2}\!\! \left(2\pi \right)^4 
\delta^4 \left( p' +k_\varphi - p - k\right) \nonumber \\
 & & \times \left[ \bar{u}'(p') \sigma_{\mu \nu} u(p)\right] \left(k^\mu e^\nu - k^\nu e^\mu \right) \,,
\end{eqnarray}
}
\ni where $k^2_A = (k_\varphi - k)^2$.

Making use of the formula for differential cross-section

\begin{equation}
\frac{d^6 \sigma(\psi \gamma \rightarrow \psi \varphi_{\rm ph})}{d^3 \vec{p}\,'_p d^3 \vec{k}_\varphi} = \frac{(2\pi)^6}{v_{\rm rel}}\,\sum_{u'}\frac{1}{2}\,\sum_u \frac{1}{2}\,\sum_e\frac{|S(\psi \gamma \rightarrow \psi \varphi_{\rm ph})|^2}{(2\pi)^4 \delta^4(0)}\,,
\end{equation}

\ni we calculate in the centre-of-mass frame, where $\vec{p} + \vec{k} = 0$ and so $\vec{p}\,' + \vec{k}_\varphi = 0$, that with $e^0 = 0$ 

\begin{eqnarray}
\frac{d^6 \sigma(\psi \gamma \rightarrow \psi \varphi_{\rm ph})}{d^3 \vec{p}\,' d^3 \vec{k}_\varphi} & = &\!\!\frac{1}{v_{\rm rel}} \,\frac{1}{256\pi^2} \left( \frac{f \zeta}{k^2_A \!-\! M^2}\right)^2 \frac{1}{E' E \omega_\varphi \omega}\, \delta^4\left(p' + k_\varphi - p - k \right)  \nonumber \\
& & \!\!\times \left[\left(E + \omega\right)^2 - m^2_\psi\right] \left[\left(E + \omega\right)^2 - m^2_\psi - m^2_\varphi + k^2_A\right]\,.
\end{eqnarray}

\ni Then,

\begin{eqnarray}
\frac{d^3 \sigma(\psi \gamma \rightarrow \psi \varphi_{\rm ph})}{d^3 \vec{k}_\varphi} & = & \frac{1}{v_{\rm rel}} \,\frac{(f \zeta)^2}{256\pi^2} \frac{1}{E' E \omega_\varphi \omega}\, \delta\left(E' + \omega_\varphi - E - \omega \right)  \nonumber \\
& & \!\!\times \frac{\left[\left(E + \omega\right)^2 - m^2_\psi\right] \left[\left(E + \omega\right)^2 - m^2_\psi - m^2_\varphi + k^2_A\right]}{(k^2_A\! -\! M^2)^2}
\end{eqnarray}

\ni with $d^3 \vec{k}_\varphi = \vec{k}\,^2_\varphi d|\vec{k}_\varphi| 2\pi \sin \theta_{\vec{k}_\varphi}$ and 

\begin{equation}
\frac{d (E' + \omega_\varphi - E - \omega)}{d |\vec{k}_\varphi|} = \frac{|\vec{k}_\varphi|}{E'} + \frac{|\vec{k}_\varphi|}{\omega_\varphi} = \frac{E + \omega}{E' \omega_\varphi} |\vec{k}_\varphi|\,. 
\end{equation}

Hence, we obtain the following integral cross-section in the centre-of-mass frame (in the channel $\psi \gamma \rightarrow \psi \varphi_{\rm ph}$):

\vspace{-0.3cm}

{\baselineskip 0.9cm

\begin{eqnarray}
\sigma(\psi \gamma \rightarrow \psi \varphi_{\rm ph})\!\! &\!\!\! = \!\!\!&\!\! \int\!\!\! d^3 \vec{k}_\varphi \frac{d^3 \sigma(\psi \gamma \rightarrow \psi \varphi_{\rm ph})}{d^3 \vec{k}_\varphi} \!= \! \int^{+1}_{-1}\!\!\!\!\!\! d\cos \theta_{\vec{k}_\varphi}\,\frac{1}{v_{\rm rel}} \frac{(f \zeta)^2}{128\pi} \frac{2|\vec{k}_\varphi|}{E}\,\frac{2\omega(E \!+\! \omega) \!-\! m^2_\varphi \!+\! k^2_A}{(M^2 \!- k^2_A )^2} \nonumber \\
&\!\!\!=\!\!\! & \!\! \int^{k^2_{A+}}_{k^2_{A-}} \!\!d k^2_A \;\frac{1}{v_{\rm rel}} \;\frac{(f \zeta)^2}{128\pi}\;
\frac{1}{E\omega}\; \frac{2\omega(E + \omega) - m^2_\varphi + k^2_A}{(M^2 \!- k^2_A )^2}\nonumber \\
&\!\!\!=\!\!\! & \!\!\frac{1}{v_{\rm rel}} \frac{(f \zeta)^2}{128\pi}\frac{1}{E\omega}\!\left\{\! \frac{4\omega|\vec{k}_\varphi|\left[\!2\omega(E \!\!+\!\! \omega)\!\!-\!\!m^2_\varphi \!\!+\!\! M^2 \!\right]}{(M^2 \!\!-\!\! m^2_\varphi \!\!+\!\! 2\omega\omega_\varphi)^2 \!\!-\!\! 4\omega^2\vec{k}^{\,2}_\varphi} \!+\! \ln|1\!\! -\!\!\frac{4\omega|\vec{k}_\varphi|}{M^2 \!\!-\!\! m^2_\varphi \!\!+\!\! 2\omega(\omega_\varphi \!\!+\!\! |\vec{k}_\varphi|)}|\!\right\}, \nonumber \\
\end{eqnarray}
}

\vspace{-0.5cm}

\ni where

\vspace{-0.3cm}

\begin{equation}
k^2_A = (k_\varphi - k)^2 = m^2_\varphi - 2\omega\omega_\varphi + 2\omega |\vec{k}_\varphi|\cos \theta_{\vec{k}_\varphi}\;,
\end{equation}

\ni $d k^2_A = 2\omega |\vec{k}_\varphi|d\cos \theta_{\vec{k}_\varphi}$ and $k^2_{A\pm} = m^2_\varphi - 2\omega\omega_\varphi \pm 2\omega |\vec{k}_\varphi|$. In Eq. (16) we have

\begin{equation}
\omega_\varphi = \sqrt{\vec{k}^{\,2}_\varphi + m^2_\varphi} = \frac{(E \!+\! \omega)^2 \!-\! m^2_\psi \!+\! m^2_\varphi }{2(E + \omega)} = \omega + \frac{m^2_\varphi}{2(E + \omega)} \;.
\end{equation}

\vspace{0.1cm}

\ni due to $E' + \omega_\varphi = E + \omega$ and $\vec{p}\,' + \vec{k}_\varphi = \vec{p} + \vec{k} = 0$ (with $E' = \sqrt{\vec{k}^{\,2}_\varphi + m^2_\psi}$ , $E =  \sqrt{\vec{k}^{\,2} + m^2_\psi}$ and $\omega = |\vec{k}|$). Here, the relative velocity of sterinos and photons colliding in the centre-of-mass frame is


\begin{equation}
v_{\rm rel} = \frac{|\vec{p}|}{E} + \frac{|\vec{k}|}{\omega} = \frac{E+\omega}{E}\,.
\end{equation}

\vspace{0.1cm}

We can see that the weak photoproduction of a steron from a sterino, $\psi \gamma \rightarrow \psi \varphi_{\rm ph}$, is --- beside the processes ${\psi}\psi \rightarrow {\psi}\psi $ and $\varphi_{\rm ph}\varphi_{\rm ph} \rightarrow \gamma\gamma$ --- a fundamental interaction in our model of hidden sector, since the three processes are provided by the simple exchange of a single sterile $A$ boson in $t$-channel, mediating our new weak coupling (1). The weak interaction (1) gives, in addition, the virtual processes  $A^* \rightarrow \gamma^* $ and $\gamma^* \rightarrow A^*$ generated spontaneously by $<\!\!\varphi\!\!>_{\rm vac} \neq 0$. They can be inserted into simpler interaction diagrams as, for instance, in the case of $\varphi_{\rm ph} \rightarrow A^* \gamma \rightarrow \gamma\gamma$, where $A^* \rightarrow \gamma$. 


\vspace{0.4cm}  

{\bf 3. Decay of sterons into couples of photons}

\vspace{0.4cm}

Now, consider the weak decay of physical sterons into couples of photons, $\varphi_{\rm ph} \rightarrow A^* \gamma \rightarrow \gamma \gamma$. Such a process is provided by a part of the coupling (1),

\begin{equation}
- \frac{1}{2} \sqrt{f}\left(<\!\!\varphi\!\!>_{\rm vac} + \varphi_{\rm ph}\right) F_{\mu \nu}A^{\mu \nu}\,,
\end{equation}

\ni giving the following $S$-matrix element (in an obvious notation):

\vspace{-0.3cm}

{\baselineskip 0.9cm

\begin{eqnarray}
S(\varphi_{\rm ph} \rightarrow \gamma\gamma)&\!\! = \!\!&-i\,\frac{1}{4}\, 
\frac{f<\!\!\varphi\!\!>\!_{\rm vac} }{k^2_A \!-\! M^2}\! \left[\frac{1}{(2\pi)^9}\,
\frac{1}{8\omega_1 \omega_2 \omega_\varphi}\right]^{\!1/2}\!\!\! (2\pi)^4 \delta^4 (k_1 \!+\! k_2 \!-\! k_\varphi)  \nonumber \\ 
 & \!\! &\times (k_{1 \mu}e_{1 \nu} - k_{1 \nu} e_{1 \mu})(k^\mu_2 e^\nu_2 - k^\nu_2 e^\mu_2) \,,  
\end{eqnarray}
}
\ni where $k^2_A = k^2_1 = k^2_2 = 0$. 

From the formula for differential decay rate

\begin{equation}
\frac{d^6 \Gamma(\varphi_{\rm ph} \rightarrow \gamma\gamma)}{d^3 \vec{k}_1 d^3 \vec{k}_2} = (2\pi)^3\sum_{e_1} \sum_{e_2} \frac{|S(\varphi_{\rm ph} \rightarrow \gamma\gamma)|^2}{(2\pi)^4 \delta^4(0)}\,, 
\end{equation}

\ni we calculate at rest, where $\vec{k}_\varphi = 0$ and so $\vec{k}_1 + \vec{k}_2 = 0$ (thus $\omega_1 = \omega_2 = \omega_ \varphi/2 = m_\varphi/2$), that with $e^0_1 = 0$ and $e^0_2 = 0$

\begin{equation} 
\frac{d^6 \Gamma(\varphi_{\rm ph} \rightarrow \gamma\gamma)}{d^3 \vec{k}_1 d^3 \vec{k}_2} = \frac{1}{64\pi^2}\left(\frac{f<\!\!\varphi\!\!>\!_{\rm vac} }{M^2}\right)^{\!\!2} \omega_ \varphi \,\delta^4(k_1 \!+\! k_2 \!-\! k_\varphi)\,.
\end{equation}

\ni Then, 

\begin{equation}
\frac{d^3 \Gamma(\varphi_{\rm ph} \rightarrow \gamma\gamma)}{d^3 \vec{k}_1} = \frac{1}{64\pi^2} \left(\frac{f<\!\!\varphi\!\!>\!_{\rm vac}}{M^2}\right)^{\!\!2} \omega_\varphi \,\delta(\omega_1 + \omega_2 - \omega_\varphi)
\end{equation}

\ni with $d^3 \vec{k}_1 = \vec{k}_1^{\,2} d|\vec{k}_1| 2\pi \sin \theta_{\vec{k}_1} d \theta_{\vec{k}_1}$ (leading to $\vec{k}_1^2 d|\vec{k}_1| 4\pi)$ and

\begin{equation} 
\frac{d(\omega_1 + \omega_2 - \omega_\varphi)}{d|\vec{k}_1|} = \frac{|\vec{k}_1|}{\omega_1} + \frac{|\vec{k}_1| }{\omega_2}  = 2 \frac{|\vec{k}_1| }{\omega_1} = 2\;.
\end{equation}

\vspace{0.2cm}

Hence, we obtain the following integral decay rate at rest (in the channel $\varphi_{\rm ph} \rightarrow \gamma\gamma$):

\begin{equation}
\Gamma(\varphi_{\rm ph} \rightarrow \gamma\gamma) = \frac{1}{2} \int d^3 \vec{k}_1\frac{d^3 \Gamma(\varphi_{\rm ph} \rightarrow \gamma\gamma)}{d^3 \vec{k}_1} = \frac{1}{256\pi} \left(\frac{f<\!\!\varphi\!\!>\!_{\rm vac} }{M^2}\right)^{\!\!2} \omega^3_\varphi \;,
\end{equation}

\vspace{0.5cm}

\ni where we have

\begin{equation} 
\omega_\varphi = m_\varphi \,.
\end{equation}

Stable sterinos are required to form in our model the thermal cold dark matter. Then [5],

\begin{equation}
<\!\!\sigma_{\rm ann}(\bar{\psi}\psi)v_{\rm rel}\!\!> \sim {\rm pb} \;\;{\rm when} \;\; \Omega_{\rm DM}\, h^2 \sim 0.11
\end{equation}

\ni (${\rm pb} = 10^{-36} {\rm cm}^2$). In this case, we put


\begin{equation}
\sigma_{\rm ann}(\bar{\psi}\psi) \sim \sigma(\bar{\psi}\psi \rightarrow \varphi_{\rm ph} \gamma) + \sum_f \sigma(\bar{\psi}\psi \rightarrow \bar{f}f) \,,
\end{equation}

\ni where $f$ are charged leptons and all quarks, except only for top if $m_t > m_\psi$ (masses of charged leptons and active quarks are neglected {\it versus} $E_\psi \sim m_\psi$). Specifically, assuming in the total annihilation cross-section (29) that 

\begin{equation} 
m_\psi^2 \sim m_\varphi^2 \sim (10^{-3}\;{\rm to}\; 1)<\!\!\varphi\!\!>\!_{\rm vac}^{\!\!2}\;\;,\;\; M^2 \sim \,<\!\!\varphi\!\!>\!_{\rm vac}^{\!\!2}  
\end{equation}

\ni and putting boldly that 

\begin{equation} 
f \sim e^2 \simeq 0.917 \;\;,\;\;\zeta \sim 1 \,, 
\end{equation}

\ni we can estimate [2] from the experimental value (28) of $\Omega_{\rm DM}\,h^2$ the masses


\begin{equation} 
m_\psi \sim m_\varphi \sim (13\;{\rm to}\; 770)\, {\rm GeV}\;\;,\;\; M \sim (400\;{\rm to}\; 770)\, {\rm GeV}\;,  
\end{equation} 

\ni respectively.

With this tentative estimation of $m_\varphi$, we calculate from Eq. (26) the following steron integral decay rate at rest (in the channel $\varphi_{\rm ph} \rightarrow \gamma\gamma$):

\begin{equation} 
\Gamma(\varphi_{\rm ph} \rightarrow \gamma\gamma) \sim (1.4\times 10^{-4}\;{\rm to}\; 8.1)\, {\rm MeV} \;,
\end{equation}

\ni respectively.

If a steron decays really at rest, then the two resulting photons have energy $m_\varphi/2$ each.

\vfill\eject

\vspace{0.4cm}

\ni {\bf 4. Conclusions} 

\vspace{0.4cm}

Due to the photonic portal open in our model to the hidden sector of the Universe, the resulting weak interactions between the cold dark matter and the \SMo sector get a magnetic structure proportional to a weak dark-matter magnetic moment generated spontaneously by $<\!\!\varphi\!\!>\!\!_{\rm vac} \!\neq\! 0$ ({\it cf.} Ref. $\!$[4] for scattering of nucleons on~dark matter; {\it cf}. also Refs. [6] and [7] proposing concepts of "magnetic inelastic dark matter"\,and "$\!\,$luminous dark matter", respectively). In the present note, we discussed weak interactions induced by the photonic portal inside the hidden sector, focusing on a new possible astrophysical process of photoproduction of sterile scalars (sterons) from sterile Dirac fermions (sterinos) {\it i.e.}, from the cold dark matter.

Sterile scalars, weakly photoproduced from cold dark matter and decaying subsequently into couples of gamma rays, may be an important source for some secondary gamma rays appearing in astrophysics. In such a case, primary gamma rays, if energetic enough, may create, in sufficiently dense configurations of dark-matter sterinos, a new astrophysical phenomenon of secondary-gamma-ray cascades. Primary gamma rays, probing the cold dark matter, may be provided {\it e.g.} by gamma bursts. For an analogy, see the discussion in Ref. [8] on possible probing the cold dark matter with particle jets emitted in active galactic nuclei.

\vfill\eject

\vspace{1.0cm}

{\centerline{\bf References}}

\vspace{0.4cm}

\baselineskip 0.7cm

{\everypar={\hangindent=0.65truecm}
\parindent=0pt\frenchspacing

{\everypar={\hangindent=0.65truecm}
\parindent=0pt\frenchspacing

~[1]~W.~Kr\'{o}likowski, {\it Acta Phys. Polon.} {\bf B 39}, 1881 (2008); arXiv: 0803.2977 [{\tt hep--ph}]; {\it Acta Phys. Polon.} {\bf B 40}, 111 (2009); arXiv: 0903.5163 [{\tt hep--ph}]; {\it Acta Phys. Polon.} {\bf B 40}, 2767 (2009).

\vspace{0.2cm}

~[2]~W.~Kr\'{o}likowski, {\it Acta Phys. Polon.} {\bf B 41}, 1277 (2010).

\vspace{0.2cm}

~[3]~{\it Cf. e.g.}  J. March-Russell, S.M. West, D. Cumberbath and D.~Hooper, {\it J. High Energy Phys.} {\bf 0807}, 058 (2008).

\vspace{0.2cm}

~[4]~W.~Kr\'{o}likowski, arXiv: 1004.1048 [{\tt physics.gen--ph}]. 

\vspace{0.2cm}

~[5]~C. Amsler {\it et al.} (Particle Data Group), {\it Review of Particle Physics}, {\it Phys. Lett.} {\bf B 667}, 1 (2008). 

\vspace{0.2cm}

~[6]~S.~Chang, N.~Weiner and I.~Yavin, arXiv: 1007.4200 [{\tt hep--ph}]. 

\vspace{0.2cm}

~[7]~B.~Feldstein, P.W.~Graham and S.~Rajendran, arXiv: 1008.1988  [{\tt hep--ph}]. 

\vspace{0.2cm}

~[8]~M.~Gorchtein, S.~Profumo and L.~Ubaldi, arXiv: 1008.223  [{\tt astro--ph.HE}]; and references therein.

\vspace{0.2cm}

\vfill\eject

\end{document}